\documentclass{aa}
\usepackage{graphicx}
\usepackage{natbib}
\usepackage{txfonts}
\bibpunct{(}{)}{;}{a}{}{,}    
\usepackage{cancel}
\usepackage{xcolor}
\usepackage[colorlinks=true,citecolor=blue,linkcolor=blue]{hyperref}

\def\apjs{ApJS}

\def\aap{Astron. Astrophys.}

\begin{document} 
\title{Power spectra of solar brightness variations at different inclinations}
\author{N.-E. Nèmec,
          \inst{1}
          A. I. Shapiro\inst{1},
          N. A. Krivova\inst{1},
           S. K. Solanki \inst{1,2},
          R. V. Tagirov\inst{1,3}, 
          R. H. Cameron \inst{1},
           and S. Dreizler \inst{4}
          }

   \institute{Max-Planck-Institut für Sonnensystemforschung, Justus-von-Liebig-Weg 3, D-37077 Göttingen, Germany\\
              \email{nemec@mps.mpg.de} 
               \and School of Space Research, Kyung Hee University, Yongin, Gyeonggi, 446-701, Korea
               \and Astrophysics Group, Imperial College London, London SW7 2AZ, UK
         \and Institut für Astrophysik, Georg-August-Universität Göttingen, Friedrich-Hund-Platz 1, D-37077 Göttingen, Germany           
             }

   \date{\today}

 
  \abstract
   {Magnetic features on the surfaces of cool stars cause variations of their brightness. Such variations have been extensively studied for the Sun.  Recent planet-hunting space telescopes allowed measuring brightness variations in hundred thousands of other stars. The new data posed the question of how typical is the Sun as a variable star. Putting solar variability into the stellar context suffers, however, from the bias of solar observations being made from its near-equatorial plane, whereas stars are observed at all possible inclinations.}
   {We model solar brightness variations at timescales from days to years as they would be observed at different inclinations. In particular, we consider the effect of the inclination on the power spectrum of solar brightness variations. The variations are calculated in several passbands  routinely used for stellar measurements. }
   {We employ the Surface Flux Transport Model (SFTM) to simulate the time-dependent spatial distribution of magnetic features on both near- and far-sides of the Sun. This distribution is then used to calculate solar  brightness variations following the SATIRE (Spectral And Total Irradiance REconstruction) approach.}
 {We have quantified the effect of the inclination on solar brightness variability at timescales down to a day. Thus, our results allow making solar brightness records directly comparable to those obtained by the planet-hunting space telescopes. Furthermore, we decompose solar brightness variations into the components originating from the solar rotation and from the evolution of magnetic features.}
   {}
  

\titlerunning{Power spectra of solar variability}
\authorrunning{Nèmec et al.}
\maketitle
 
%

\section{Introduction}

Recent planet-hunting missions such as CNES' CoRoT \citep[Convection, Rotation and planetary Transit,][]{COROT,COROT2}, NASA's \textit{Kepler} \citep{KEPLER} and the Transiting Exoplanet Survey Satellite  \citep[TESS,][]{TESS} have opened up new possibilities for studying stellar variability up to timescales of the rotational period and in some cases beyond. \citep{Timo2017,Montet2017}. 
A plethora of data obtained by these missions underlines the needs for a better understanding and modelling of stellar brightness variations. 
One of the possible approaches for such modelling is to rely on the solar paradigm, i.e. to take a model which reproduces observed variability of solar brightness and extend it to other stars. For example, such an approach has been used by \cite{witzkeetal2018} who extended the Spectral And Total Irradiance REconstruction  \cite[SATIRE,][]{Fligge2000,Krivova2003} model of solar brightness variability to calculate brightness variations over the timescale of the activity cycle in stars with different metallicities and effective temperatures. 
 Later,  \cite{Veronika_rot} utilised a similar model to investigate how the amplitude of the rotational stellar brightness variability as well as the detectability of stellar rotation periods depend on the metallicity.  Here, we perform one more extension of the SATIRE model to study how the amplitude of solar brightness variability depends on the angle between solar rotation axis and directions to the observer (hereafter, inclination). 

The brightness variability of the Sun is brought about by the magnetic features (such as dark spots and bright faculae) on its surface \citep[see, e.g. reviews by][]{TOSCA2013, MPS_AA}. The visibility of the magnetic features and their brightness contrasts depend on the position of the observer relative to the solar rotation axis. This causes the solar brightness variability to depend on the  inclination. A quantitative assessment of such a dependence is of particular importance for answering the question of how solar photometric variability compares to that of other stars. To properly address this question one needs to take into account that the Sun is observed from its near-equatorial plane (i.e. at inclinations close to $90^{\circ}$), while stars are observed at random, mostly unknown, inclinations. 

The effect of the inclination on solar variability can only be assessed with models since
solar brightness has never been measured out of ecliptic. For example, to account for possible long-term climate response to the change of the Earth’s orbital inclination in relation to solar equator, \cite{Vieira2012}  developed a model based on combining synoptic maps and disk images obtained from the Helioseismic and Magnetic Imager \citep[HMI,][]{HMI} data. They found  that {\it on timescales of several thousand years} the total solar irradiance (TSI) variability due to the change of the  Earth’s orbital inclination is negligibly small.

A number of studies have modelled the dependence of solar brightness variability on the inclination {\it over the timescale of the 11-year activity cycle}. These studies have been motivated by ground-based observations of Sun-like stars that revealed that the Sun exhibits lower photometric variability {\it on the activity cycle timescale} than most Sun-like  stars with near-solar levels of magnetic activity \citep[][]{Lockwood1990,Lockwood2007,Radick2018}.  \cite{Schatten1993} proposed that this enigmatic behaviour of the Sun is due to its equator-on view from the Earth. He found that the amplitude of the activity cycle in solar brightness significantly increases with decreasing inclination.
Later, \cite{Knaack2001} and \cite{Shapiro2014} employed a more accurate model and also found an increase of the variability for the out-of-ecliptic observer, but the effect of the inclination appeared to be considerably weaker than that reported by \cite{Schatten1993}. All in all, the current consensus is that the effect of inclination cannot explain the low variability of the Sun {\it on the activity cycle timescale} and, consequently, other explanations have been proposed \citep{Shapiro2016,witzkeetal2018, Karoff2018}.

\cite{Schatten1993},\cite{Knaack2001} and \cite{Shapiro2014} assumed an axisymmetric band-like distribution of faculae and spots. Such an assumption is justifiable for modelling solar brightness variations {\it on the activity cycle timescale} but it does not allow modelling brightness variability {\it on the solar rotational timescale}. Indeed, the activity cycle variability is caused by the overall modulation in the solar surface coverage by magnetic features from activity minimum to maximum and depends only on the time-averaged surface distribution of magnetic features (which can be approximated by the axisymmetric band-like structure rather well). In contrast, rotational variability is caused by the evolution of individual magnetic features and their transits across the visible solar disc as the Sun rotates. Consequently, it depends on the exact distribution of magnetic features.

An attempt to model the effect of the inclination on the rotational solar brightness variability has been recently made by \cite{Shapiro2016}. They  used distribution of magnetic features on the visible solar disk provided by \cite{Yeo2014} and obtained the distribution of magnetic features on the far-side of the Sun (part of which would become visible for the observer not bound to the Earth) assuming that the near- and far-sides of the Sun are point-symmetric with respect to each other through the centre of the Sun.  They found that an observer bound to the ecliptic plane witnesses the Sun to be spot-dominated {\it on the rotational timescale}, but with decreasing inclination the amplitude of the rotational variability decreases (in contrast to the brightness variability {\it on the activity timescale} which increases with decreasing inclination) and the facular contribution becomes dominant. Despite being more advanced relative to previous studies, the assumption of the point-symmetric distribution of solar magnetic features employed in \cite{Shapiro2016}  did not allow accounting for the appearance and disappearance of magnetic features which rotate in and out of the visible
solar disc. This led to a number of artefacts which did not allow  studying the effects of the inclination on the detectability of stellar rotation periods. These effects might play, however, an important role in understanding the observed distribution of rotation periods in Kepler stars \citep{Reinhold2019, VanSaders2019}. Also these artefacts hindered the accurate assessment of the inclination effect on the timescale of solar rotation.  Such an assessment is, in turn, needed for the interpretation of the data from the planet-hunting missions. For example, the \textit{Kepler} data indicated that also solar brightness variability {\it on the timescale of solar rotation} appears to be lower than that in most of the stars with known near-solar fundamental parameters and rotation periods \cite{Timo2020}. 

Here we take a different approach from \cite{Shapiro2016} and utilise a surface flux transport model \citep[SFTM,][]{Cameron2010} to obtain the distribution of solar magnetic features over the entire solar surface (i.e. on both near- and far-sides of the Sun).  This distribution is then fed into the SATIRE model to calculate the solar brightness variability  for different solar activity levels, various photometric filter system used in stellar observations, and at different inclinations. In particular, we show how the change of the inclination affects the power spectrum of solar brightness variations. This allows studying the impact of the inclination on brightness variability depending on the timescale of the variability. 
In Sect.~\ref{Methods} we describe how we compute the solar disc area coverages by magnetic features from the SFTM and then calculate the brightness variations following the SATIRE model.  We also list the main parameters of the model and explore their impact on the brightness variations.In Sect.~\ref{ecliptic} we show how the strength of an individual cycle affects the solar photometric variability in different passbands, before we move to different inclinations in Sect.~\ref{inclination}. In Sect.~\ref{inclination} we also decompose the solar brightness variability into components arising from the evolution of magnetic features and from the solar rotation. We present our main conclusions in Sect.~\ref{conclusion}.

\section{Methods}\label{Methods}

\subsection{Calculating brightness variations}\label{sec:SATIRE}

We build our method on the SATIRE model, in which  brightness variations on timescales longer than a day are attributed to the emergence and evolution of magnetic field on the surface of the Sun, as well as on solar rotation \citep[][]{Fligge2000,Krivova2003}.
The photospheric magnetic features are divided into three main classes: sunspot umbra ($u$), sunspot penumbra ($p$), and faculae ($f$). 
The intensities of these features and that of the quiet Sun ($q$) depend on the wavelength and the cosine of heliocentric angle $\theta$ ($\mu = \mathrm{cos}\theta$), but are time-independent. The intensities were computed
by \cite{Unruh1999} \citep[following][]{Castelli1994} with the use of the spectral synthesis
code ATLAS9 \citep{Kurucz1992}.
 The 1D atmospheric structures of umbra, penumbra, and quiet Sun were calculated using radiative equilibrium models, while the facular model is a modified version of FAL-P by \cite{Fontenla1993}.

The Spectral Solar Irradiance $S(t,\lambda_w)$ (i.e. spectral radiative flux from the Sun, normalized to one AU), where $t$ is the time and $\lambda_w$ the wavelength (not to be confused with $\lambda$ used for the latitude later in this paper), is calculated by summing up the intensities weighted by the corresponding fractional disc area coverages of the magnetic features (designated with the index $k$) as given by

\begin{equation}
    S(t,\lambda_w) = S^{q}(\lambda_w) +\sum_{mn} \sum_{k}(I_{mn}^{k}(\lambda_w)-I_{mn}^{q}(\lambda_w)) \, \alpha_{mn}^{k}(t)
    \Delta\Omega_{mn}.
\label{SSI}
\end{equation}

\noindent Here the summation is done over the pixels of the magnetograms and the $m$ and $n$ indexes are the pixel coordinates (longitude and latitude, respectively), $\alpha_{mn}^{k}$ is the fraction of pixel ($m$,$n$) covered by magnetic feature $k$, $\Delta \Omega_{mn}$ is the solid angle of the area on the solar  disc corresponding to one pixel, as seen from the distance of 1 AU, and S$^{q}$ is the quiet Sun irradiance, defined as

\begin{equation}
   S^{q}(\lambda_w) = \sum_{mn} I_{mn}^{q}(\lambda_w)\Delta\Omega_{mn}.
\end{equation}

\noindent 
The solid angles of pixels as well as corresponding intensity values depend on the vantage point of the observer. Consequently, the solar irradiance values $S(t,\lambda_w)$ given by Eq. (\ref{SSI}) also depend on the vantage point of the observer and, in particular, on the inclination.

\subsection{Surface flux transport model}\label{sec:SFTM}

To simulate the full surface distribution of magnetic features, we use the SFTM in the form presented in \cite{Cameron2010}. The SFTM describes the passive transport of the radial component of the magnetic field B, considering the effects of differential rotation $\Omega(\lambda)$ (with $\lambda$ being the latitude), meridional flow $\nu(\lambda)$ at the solar surface, and a horizontal surface diffusion  thanks to a non-zero diffusivity $\eta_H$. The emerged active regions gradually disperse due to the radial diffusion $\eta_r$, with the flux finally decaying after cancellation between opposite polarities, where they overlap. The governing equation is

\begin{equation}
\begin{split}
  \frac{\partial B}{\partial t} &= - \Omega(\lambda)\frac{\partial B}{\partial \phi} - \frac{1}{R_{\odot}\cos\lambda}\frac{\partial}{\partial \lambda}(\nu(\lambda)B \cos(\lambda))\\&+\eta_H\left ( \frac{1}{R_{\odot}^{2}\cos\lambda}\frac{\partial}{\partial \lambda}\left ( \cos(\lambda)\frac{\partial B}{\partial \lambda} \right ) + \frac{1}{R_{\odot}^{2}\cos^{2}\lambda} \frac{\partial ^{2}B}{\partial\phi^{2}}\right ) \\ &+D(\eta_r) + S(\lambda,\phi,t),
\end{split}
\label{flux}
\end{equation}

\noindent where $R_{\odot}$ is the solar radius, $\phi$ is the longitude of the 
active region, and $D$ is a linear operator that describes the decay due to radial
diffusion with the radial surface diffusivity $\eta_r$. For the linear operator $D$ the form of \cite{Baumann2006} was used. 
The horizontal diffusivity $\eta_H$ was taken to be 250 km$^{2}$s$^{-1}$ as in \cite{Cameron2010}  and the radial surface diffusivity $\eta_r$ was set to  25 km$^{2}$s$^{-1}$ according to \cite{Jiang2011_2}. The time average (synodic) differential rotation profile was taken from \cite{Snodgrass1983} and is given as (in degree per day):
\begin{equation}
\Omega(\lambda) = 13.38 -2.3 \cdot \sin^{2}\lambda -1.62\cdot \sin^{4}\lambda. 
\label{diff_rot}
\end{equation}

 The time-averaged meridional flow is expressed following \cite{vanBallegooijen1998}, namely,
\begin{equation}
\nu(\lambda)=
  \begin{cases}
    11\cdot \sin(2.4\lambda) \quad \text{m/s,} & \text{where} \quad \lambda \leq 75 ^\circ \\
	0,         & \text{otherwise}.
  \end{cases}
 \label{merflow}
\end{equation}

The source term $S(\lambda,\Phi,t)$ in Eq. (\ref{flux}) describes the magnetic flux, which is prescribed to be in the form of  two patches with opposite polarities \citep[][]{vanBallegooijen1998,Baumann2004}. The patches are centred at $\lambda_{+}$ and $\phi_{+}$ for the positive polarity patch and $\lambda_{-}$ and $\phi_{-}$ for the negative polarity patch. The field of each patch is given by

\begin{equation}
B^{\pm}(\lambda,\phi) = B_{\mathrm{max}}\left(\frac{0.4 \Delta \beta}{\delta}\right)^{2} e^{-2[1-\cos(\beta_{\pm}(\lambda,\phi))]/\delta^{2}} ,
\label{source}
\end{equation}

\noindent where $B^{\pm}$ is the flux density of the positive and negative polarity,
$\beta_{\pm}$($\lambda$,$\phi$) are the heliocentric angles between point ($\lambda$, $\phi$) and the centres of the polarity patches, $\Delta \beta$ is the separation between the two polarities and $\delta$ is the size of the individual polarity patches, taken to be 4$^\circ$. $B_{\mathrm{max}}$ is a scaling factor introduced by \cite{Cameron2010} and \cite{Jiang2011_2} and was fixed to 374 G. This value was found by forcing the total unsigned flux to match the measurements from the Mount Wilson and Wilcox Solar Observatories.

\cite{Jiang2011_1} constructed a semi-empirical source term $S(\lambda,\Phi,t)$ for the 1700--2010 period so that its statistical properties reflect those of the Royal Greenwich Observatory sunspot record. Here we adopt the $S(\lambda,\Phi,t)$ term from \cite{Jiang2011_1}  but with one important modification. As an observer stationed at a vantage point outside the ecliptic sees both the near- and far-sides of the Sun (as defined by the Earth-bound observer), it is crucial to avoid any systematic differences between the active region distributions on the two sides. To this purpose we have modified  $S(\lambda,\Phi,t)$  so that the emergence of active regions happens at random longitudes, whereas the butterfly-like shape of their latitudinal emergence, as well as the number of emergences and the tilt-angle distributions, over the course of the cycle is preserved.

All in all, the adapted source term describes the emergence of active regions on the solar surface in a  statistical way. We stress that the goal of this study is not to reproduce the exact solar light curve as it would be seen from outside the ecliptic, but to study the effect of the inclination on the power spectrum of solar brightness variations at different levels of solar activity. The statistical representation of the source term is fully sufficient for this purpose.

\subsection{From magnetic fluxes to area coverages}\label{sec:ff}

The SFTM returns simulated magnetograms, with a pixel-size of 1$^\circ \rm \times$ 1$^\circ$. We follow the approach of \cite{DasiEspuig2014} and divide each pixel ($m$,$n$) into 100 sub-pixels, with a size of 0.1$^\circ \rm \times$ 0.1$^\circ$ each.

To calculate the brightness variations, we need to distinguish between spots and faculae.
The spot areas and positions at the day of emergence have been provided by \cite{Jiang2011_1} together with the source term $S(\lambda,\Phi,t)$. After spots  emerge, their positions  on the solar surface are affected by the differential rotation described by Eq.~(\ref{diff_rot}) and the meridional flow described by Eq.~(\ref{merflow}). The spot sizes are calculated by following a decay law during their evolution.
In the literature we found studies that support linear and parabolic decay laws and different values for the decay rate \citep[][]{MorenoInsertis1988, MartinezPillet1993,Petrovay1997,Baumann2005,Hathaway2008}. As \cite{Baumann2005} found, it is not possible to distinguish between a linear and parabolic decay law from, e.g., the area distribution of sunspots.
 For simplicity, we chose a linear decay law of:
\begin{equation}
    A(t) = A_0 - R_d \cdot (t - t_0),
    \label{eq:spot_decay}
\end{equation}
\noindent where $A(t)$ is the area on a given day $t$ and $t_0$ is the day on which the spot has its maximum area $A_0$  (provided in the input). The decay rate $R_d$ is measured in microsemi-hemispheres (MSH) per day and is a semi-free parameter of the model, which will be discussed in more detail in Sect. \ref{sec:params}. The decay rate $R_d$ has been studied extensively before. In particular \cite{MartinezPillet1993} have reported several values of the decay rate, ranging from 25 to 47 MSH day$^{-1}$. The value we found to be the best for our model is 80 MSH day$^{-1}$ (see detailed description of the procedure used to determine $R_d$ in Sect.~\ref{sec:params}). The slightly higher than observational estimates value obtained for our modelling can be explained by the low spatial resolution of the source term in Eq.~(\ref{source}). A group of spots might be represented by one large spot (due to the resolution of the source term), which then will decay with the rate equal to the sum of decay rates of the individual spots.

Having the spatial and temporal spot distribution, we can now correct the simulated magnetograms for the spot magnetic flux, which is important for the masking of the faculae. The correction is done on the original 1$^\circ \rm \times$ 1$^\circ$ grid corresponding to the SFTM output since in contrast to the spot distribution which is calculated on the 0.1$^\circ \rm \times$ 0.1$^\circ$ grid, we calculate more diffuse facular distribution on the original grid. If a  1$^\circ \rm \times$ 1$^\circ$ pixel is found to be free of spots the correction is equal to 0 and the magnetic field in the pixel is directly taken from the SFTM.  If a given pixel is found to be partially covered by spots the magnetic field in the pixel is corrected as

\begin{equation}
B'_{\mathrm{(m,n)}} = B_{\mathrm{m,n}} - B_{\mathrm{spot}} \cdot a^s_{\mathrm{m,n}},
\end{equation}

\noindent where $B_{\mathrm{m,n}}$ is the pixel field returned by the SFTM, $B_{\mathrm{spot}}$ is the mean magnetic field of a spot, and $a^s_{\rm mn}$ is the fractional coverage of the pixel (m,n) by spots.
The value of $B_{\rm spot}$ is taken from observations. \cite{Keppens1996} have measured the umbral and penumbral field strength of solar sunspots. We do not distinguish between umbral and penumbral regions and use an area weighted average of the values of 800 G reported in \cite{Keppens1996}.

The remaining magnetic field $B'_{\mathrm{(m,n)}}$ (with $B'_{\mathrm{(m,n)}}=B_{\mathrm{(m,n)}}$ for pixels free of spots) is then attributed to faculae and is calculated following the SATIRE approach:
\begin{equation}
\alpha^f_{\rm m,n}= 
  \begin{cases}
    \frac{B'_{\rm m,n}}{B_{\rm sat}} & \text{if B$_{\mathrm{mn}}$ $<$  B$_{\mathrm{sat}}$} \\
	1         & \text{if B$_{\mathrm{mn}}$ $\ge$ B$_{\mathrm{sat}}$},
  \end{cases}
 \label{eq:fac_f_f}
\end{equation}
\noindent where B$_{\mathrm{sat}}$ is the saturation threshold, in accordance to the SATIRE-S model \citep{Krivova2003,Wenzler2004,Ball2012}. In this model, the facular filling factor increases linearly with the magnetic field strength, eventually reaching unity at a saturation.
Given that the SFTM provides information only at time of the maximum area and during the subsequent decay of the active regions, we need to additionally consider the growth phase of the spots (i.e. take into account that they do not emerge instantaneously). We employ a linear growth law with a constant rate $R_g$ similar to the decay law given by Eq.~(\ref{eq:spot_decay}). For $R_g$ we have not found any appropriate studies so that it is treated as  a free parameter (see next section).

\subsection{Model parameters}\label{sec:params}

\begin{figure*}
\includegraphics[width=\textwidth]{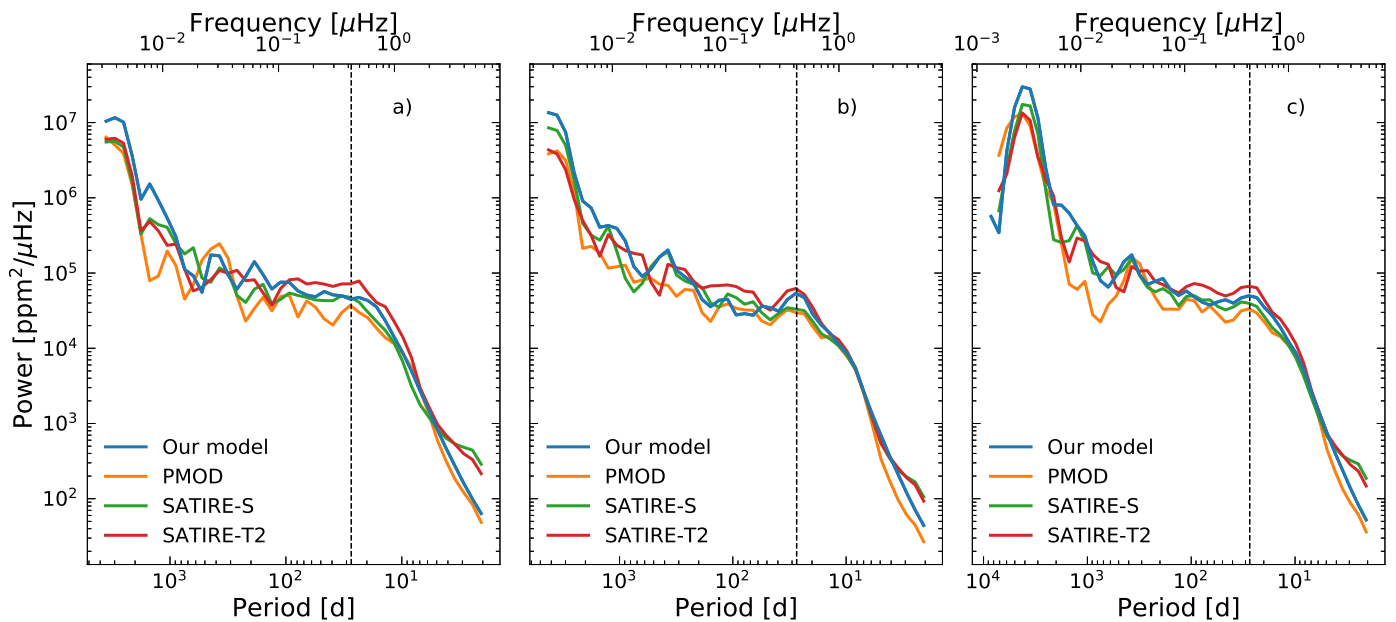}
\caption{Comparison of the power spectra of solar brightness variations produced by our model to those given by the PMOD-composite as well as SATIRE-S and SATIRE-T2 models for cycle 22 (panel a) and cycle 23 (panel b) and the combined timeseries (panel c). The vertical dashed black line indicates the synodic solar rotation period of 27.3 days.} 
\label{fig:Comparison_model_data}
\end{figure*}

\begin{table}
\caption{List of the parameters used in our model}              
\label{table:param_descricption}      
\centering                                      
\begin{tabular}{|l| l| l|}          
\hline\hline                        
Parameters & Description & Best value \\    
\hline                                   
    R$_d$ & decay rate spots  & 80 MSH day$^{-1}$\\      
    R$_g$ & growth rate spots & 600 MSH day$^{-1}$ \\
    B$_{sat}$ & saturation threshold faculae & 500 G\\
\hline                                             
\end{tabular}
\end{table}

To find the best set of model parameters, we compare power spectra of the computed TSI time series to power spectra of TSI from other sources. We use the Physikalisch-Meterologisches Observatorium Davos (PMOD) composite \citep[][version 42\_65\_1709, \url{ftp://ftp.pmodwrc.ch/pub/data}]{Froehlich2006}, which provides TSI measurements over several decades. We also use the TSI output from the SATIRE-S \citep{Yeo2014} and SATIRE-T2 \citep{DasiEspuig2016} solar irradiance variability reconstruction models. In SATIRE-S the distribution of magnetic features on the solar surface is derived from full disk images and magnetograms of the Sun, whereas in SATIRE-T2 it is derived from a SFTM but with a different source term than employed in this study.

In cycle 21, both the PMOD composite and SATIRE-S contain a significant amount of data gaps that would affect the power spectra. We therefore restrict ourselves to use cycles 22 and 23 for the determination of the best parameter set.
We show the power spectra of the solar brightness variations as presented by PMOD, SATIRE-S and SATIRE-T2 in Fig.~\ref{fig:Comparison_model_data}.
One striking difference between the datasets is that SATIRE-S and SATIRE-T2 show higher power values compared to the PMOD-composite at periods below 5 days for both considered cycles. We attribute this to aliasing effects being present in the two SATIRE-models.
Both, SATIRE-S and SATIRE-T2, give one instantaneous value of the TSI per day, whereas the PMOD-composite gives daily averages. Consequently, the difference between the power spectra appears  because of the comparison between instantaneous values (affected by aliasing) and daily averages. To avoid aliasing issue in our model output we calculate solar brightness with 6-hour cadence. We found that this leads to similar values of spectral power starting from timescales of about two days as the PMOD-composite.

We found our best set of parameters by comparing the power spectra obtained with the output of our model to those obtained with the PMOD composite. Namely, we calculated the $\chi^2$ values using the parts of the power spectra below the solar rotation period (i.e. we only considered periods shorter than 27.3 days). Despite only low-period parts of the power spectra have been used for the fit we found that we are still able to maintain a reasonable agreement on longer timescales as well. Our calculations seem to slightly overestimate the variability on the activity timescale which can be attributed to the absence of ephemeral regions in our model \citep[see discussion in][]{DasiEspuig2016}. 

\begin{figure*}[ht!]
\centering
\includegraphics[width=\textwidth]{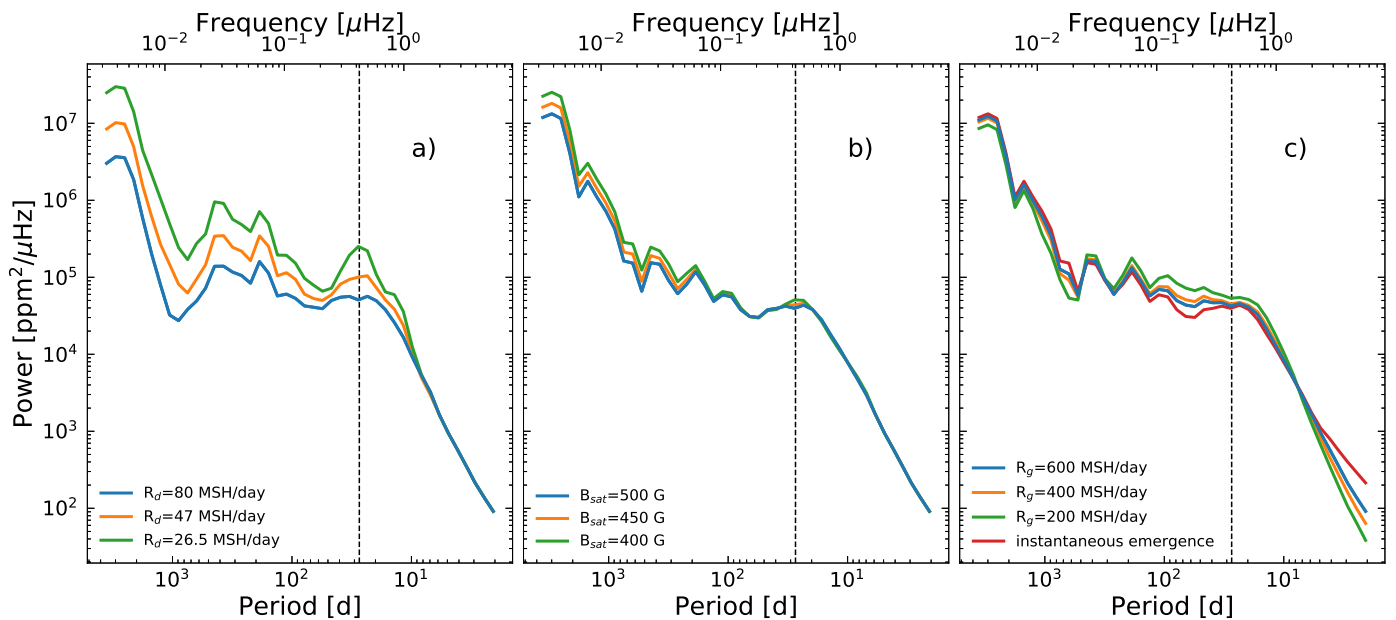}
\caption{Effect of the different parameters of the model on the brightness variations. Panel a) shows the effect of the decay rate R$_d$ on the spot component only, panel b) the effect of B$_{sat}$ on the total power spectrum and panel c) the effect of different growth rates R$_g$ on the total power spectrum compared to not having the spot growth included as depicted by the red curve. R$_d$ and R$_g$ are in units of MSH day$^{-1}$. The vertical dashed black line indicates the synodic solar rotation period at 27.3 days.} 
\label{fig:PS_params}
\end{figure*}

Let us also check how the different free parameters of our model affect the power spectrum of solar brightness variations returned by the model. The effects of the spot decay rate $R_d$  (panel a), B$_{sat}$ value (panel b), and spot growth rate $R_g$ (panel c) are illustrated in Fig.~\ref{fig:PS_params}. With decreasing spot decay rate, $R_d$, the overall area coverage of the spots is increasing, which affects  timescales longer than about 10 days \citep{Shapiro2019}. The prominent peak at the rotation period for the $R_d=26.5$ MSH day$^{-1}$ is a result of the long lifetime of the spots. The longer the spot lives, the higher the probability it reoccurs at the next rotation which leads to the formation of the rotation harmonic in the power spectrum.

The effect of the saturation threshold, $B_{sat}$, is shown in Fig.~\ref{fig:PS_params} b. We note that the facular filling factors are primarily regulated via this parameter. On the activity cycle timescale, faculae are the dominant source of variability, whereas on timescales, below 100 days, the spot component is the main driver of the variability. A value of 500 G for $B_{\rm sat}$ leads to the best fit compared to the PMOD-composite.
In contrast to the effect of the decay rate, $R_d$, the growth rate , $R_g$, shows the highest impact on timescales below 10 days (see right panel of  Fig.~\ref{fig:PS_params}). The value of 600 MSH day$^{-1}$ gives the best agreement with the PMOD composite on those timescales.

\section{Solar brightness variations as seen by an ecliptic bound observer}\label{ecliptic}

\subsection{TSI variability during activity cycles of different strengths}

\begin{figure}
\centering
\includegraphics[width=0.5\textwidth]{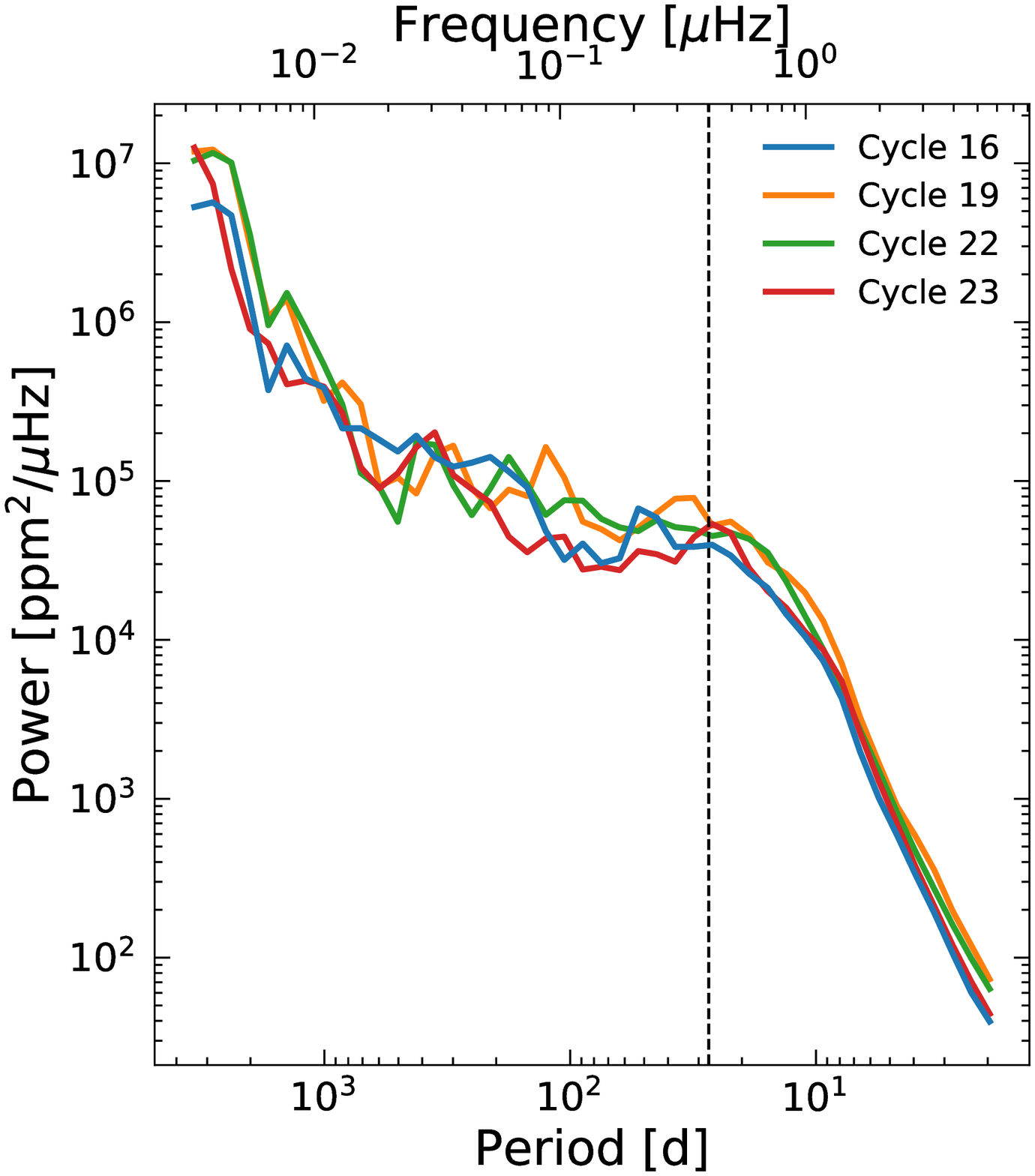}
\caption{Power spectrum of the TSI for different cycles as seen by an ecliptic bound observer. The vertical dashed black line indicates the synodic solar rotation period at 27.3 days.}
\label{fig:PS_TSI_cycles}
\end{figure}

Until now we considered the TSI variability during cycles 22 and 23. To understand the solar brightness variations in the context of stellar variability, it is important to explore different activity levels. 
With our source term we can calculate solar brightness variations back to 1700. 
In Fig.~\ref{fig:PS_TSI_cycles} we compare power spectra of the TSI variability as returned by our model for cycles 16 (one of the weakest cycle over the last 300 years), 19 (the strongest cycle observed so far), 22, and 23.
For cycle 16 and 23, a small peak at the rotation period of about 27 days can be seen.
 The profile of the power spectrum for cycle 19 is rather surprising, with two peaks on periods slightly below (25 days) 
  and above (32 days) the rotation period (see also Fig. \ref{fig:PS_filter_cycles} where the double peak structure is more easily visible). \cite{Shapiro2019} explained such a 
  double-peak structure by the cancellation of spot and facular contribution to the rotation signal. \cite{Veronika_rot} further analysed the connection between the power spectrum profile and detectability of the rotation period.

Recently a lot of effort has been put into determining stellar rotation periods from photometric observations by the \textit{Kepler} telescope \citep[see, e.g.,][]{Timo2013,McQuillan2014,Angus2018}.
Intriguingly, the detection of the rotation period of old stars with near-solar level of magnetic activity appeared to be challenging due to the low amplitude of the irradiance variability,  short lifetime of spots, and the cancellation of the rotational signal from spots and faculae \citep[][]{Aigrain2015,Shapiro2017,Reinhold2019}. In agreement with previous studies \citep[e.g.][]{Lanza_Shkolnik2014,Aigrain2015} our analysis indicates that the same star can be deemed as periodic or non-periodic  \citep[according to the definition of][]{McQuillan2014}, depending on whether it is observed at high or low activity.

\subsection{Solar variability in different passbands}

\begin{figure}
\centering
\includegraphics[width=0.5\textwidth]{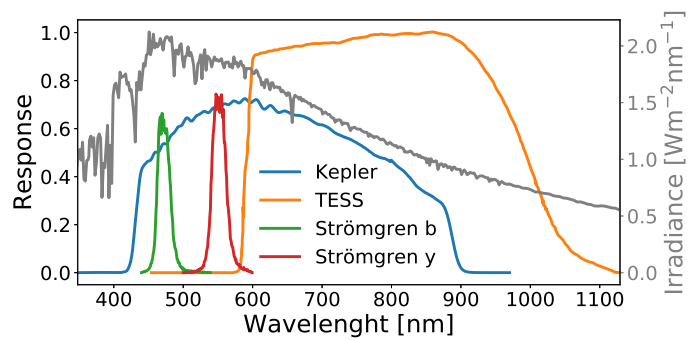}
\caption{Response functions of the different filter systems used in this work. The quiet-Sun irradiance as used by SATIRE is shown in grey.}
\label{fig:response_functions}
\end{figure}

In this section we explore solar brightness variations as they would be observed in different passbands. 
We multiply the computed spectral irradiance given by Eq.~(\ref{SSI}) with the response functions of different filter systems and then integrate over the corresponding wavelength ranges.
We consider the Strömgren filters \textit{b} and \textit{y} which have been widely used in groundbased observations to study long-term stellar photometric variability \citep{Radick2018}, as well as the \textit{Kepler} and TESS passbands. The transmission curves and the quiet-Sun spectrum (according to the SATIRE model) are shown in Fig.~\ref{fig:response_functions}. The Strömgren \textit{b} and \textit{y} filters are centred at 476 and 547 nm, respectively, so that  Strömgren \textit{b} is located around the maximum of the solar spectrum, while Strömgren  \textit{y} is shifted to the red. The primary goal of \textit{Kepler} was to find planets around solar-type  stars and its filter profile covers almost the  whole visual wavelength range. TESS is aimed at observing a large number of M dwarfs and is, consequently, more sensitive to the red part of the spectrum.

\begin{figure*}
\centering
\includegraphics[width=0.75\textwidth]{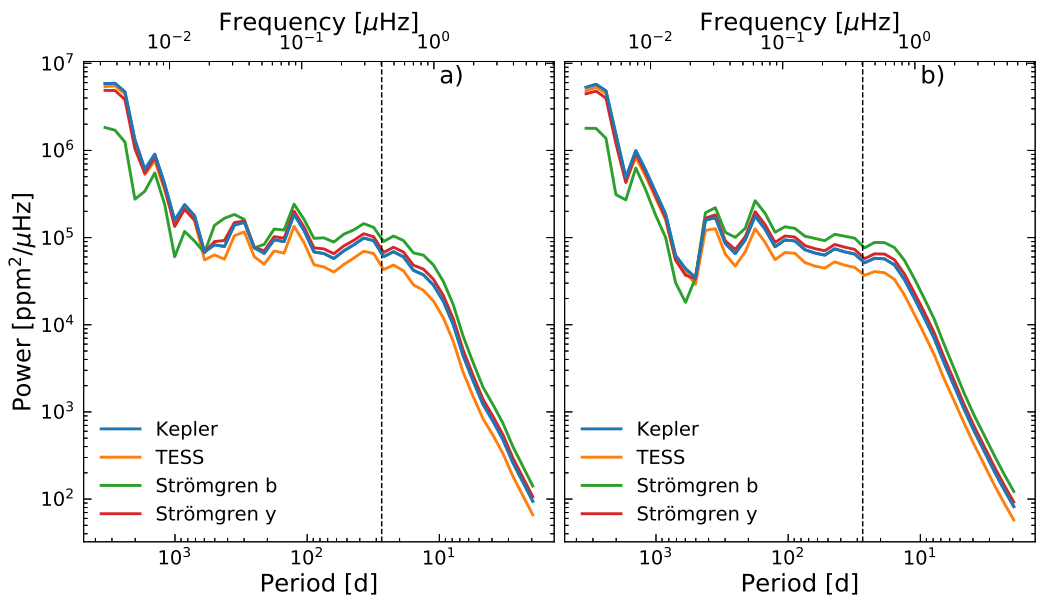}
\caption{Power spectra of solar brightness variations in different filter systems for different cycles as observed from the ecliptic.
Panel a) shows cycle 19, panel b) cycle 22. The vertical dashed black line indicates the synodic solar rotation period at 27.3 days.} 
\label{fig:PS_filter_cycles}
\end{figure*} 

We compare the different filter systems and their effect on the measured variability for different cycles as observed by a solar equator-bound observer in Fig.~\ref{fig:PS_filter_cycles}. 
Interestingly, the shapes of the power spectra are very similar on timescales below about a year.
On timescales below 1 year, the variability in the two narrow-band Strömgren filters shows the highest power,  followed by \textit{Kepler}, whereas the brightness variations as they would be observed by TESS show the lowest amplitude.

On timescales above 1 year the variability in the \textit{Kepler}, TESS and Strömgren y passband have similar strength, whereas the signal in Strömgren b is considerably lower.
 For the Strömgren b filter, \cite{Shapiro2016} have found that the facular and spot contributions to the variability almost cancel each other, hence the variability is low. The compensation is less pronounced in the other passbands.

\section{Solar brightness variations as they would be seen from out of ecliptic}\label{inclination}

In the following we refer to the inclination as the viewing angle of the observer with respect to the solar rotation axis. An inclination of 90$^\circ$ corresponds to an observer in the solar equatorial plane, while inclinations of \textless 90$^\circ$ refer to a displacement of the observer from the equatorial plane towards the North pole.

\subsection{Effect of inclination on brightness variability}\label{North}

\begin{figure*}
\centering
\includegraphics[width=0.75\textwidth]{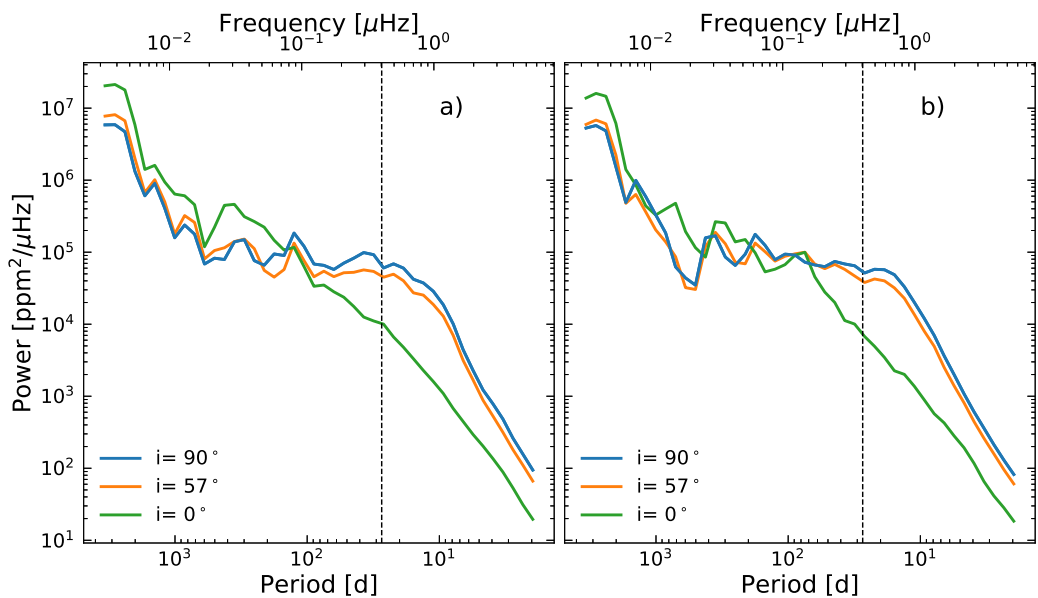}
\caption{Power spectra of solar brightness variations in the \textit{Kepler} passband with at different inclinations and two different cycles. Panel a) shows cycle 19 and b) cycle 22. The vertical dashed black lines indicate the synodic solar rotation period at 27.3 days.} 
\label{fig:PS_Kepler_TESS}
\end{figure*}

We now consider the variability during cycles 19 and 22 as it would be observed by \textit{Kepler}. The power spectra of brightness variations as they would be seen at  90$^\circ$ (i.e. from the equatorial plane), at 57$^\circ$ (which is the mean value of the inclination  for a random distribution of orientations of rotation axes), and at 0$^\circ$ (i.e. the view at the solar North pole) are plotted in Fig.~\ref{fig:PS_Kepler_TESS}.

\begin{figure*}
\centering
\includegraphics[width=0.75\textwidth]{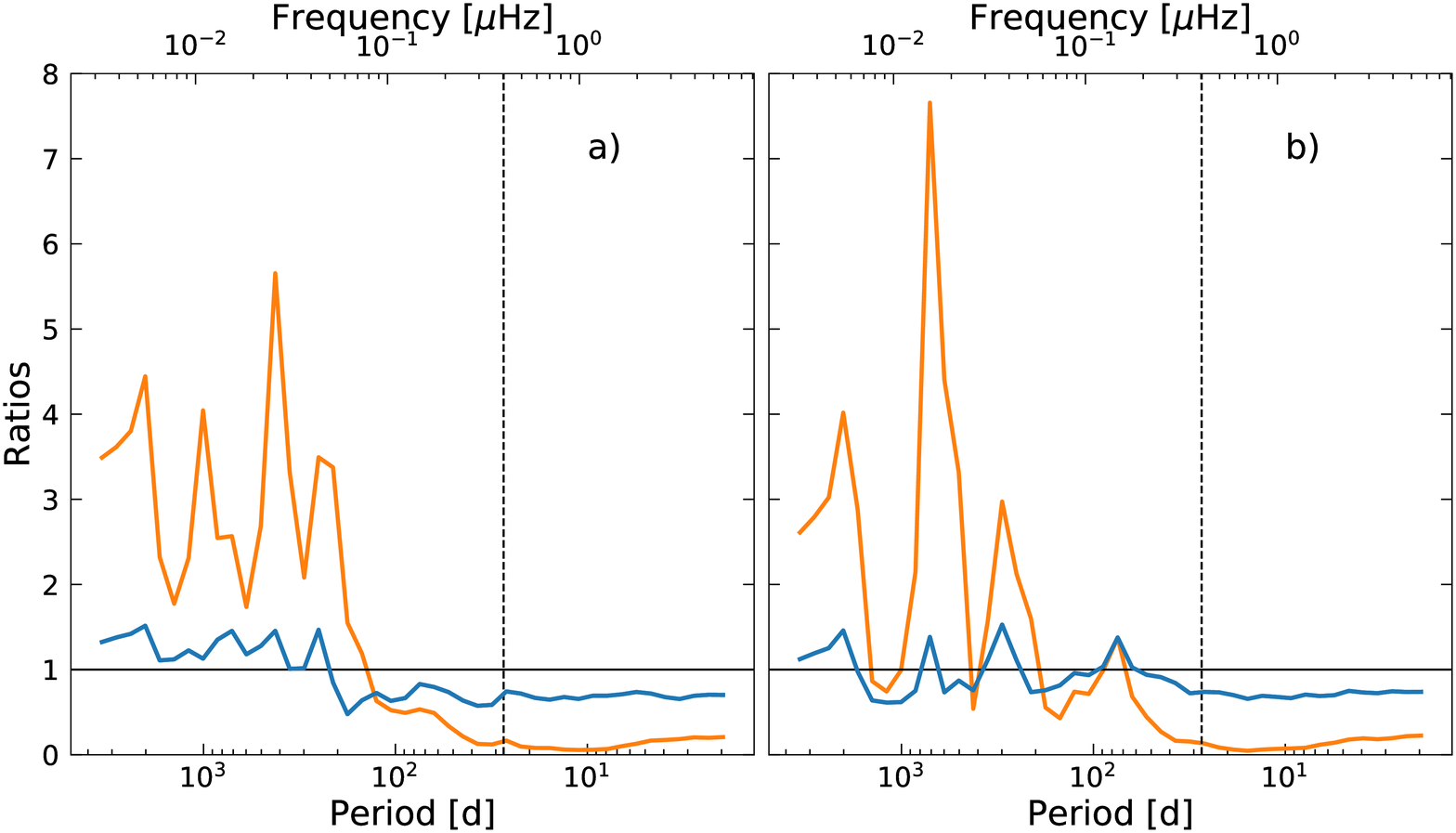}
\caption{Ratios of the power spectra of solar brightness variations for \textit{Kepler} as shown in Fig.~\ref{fig:PS_Kepler_TESS}. Blue lines represent the ratios of 57$^\circ$ to 90$^\circ$ and orange lines between 0$^\circ$ and 90$^\circ$. Panel a) shows cycle 19 and b) cycle 22, c) for cycle 22. The vertical dashed black line indicates the synodic solar rotation period at 27.3 days. The horizontal solid black line indicates a ratio of 1.} 
\label{fig:PS_incl_ratios}
\end{figure*}

The power at the rotational timescale drops with decreasing inclination, but the variability on the activity timescale increases. This effect is not strong between 90 and 57$^\circ$ inclination, but significant between 90 and 0$^\circ$.
 Interestingly, the double-peak structure of cycle 19 that has been described before for the ecliptic-bound observer, is also present for the inclination of 57$^\circ$, although the peaks are less pronounced. 
 For the observer at 0$^\circ$, the power in the signal below 100 days is significantly lower than for the 90 and 57$^\circ$ vantage point. However, on timescales longer than 100 days, the power becomes higher compared to the other vantage points. We discuss this result in more detail in Sect.~\ref{disentangle}. 
 We also show the power spectra of brightness variations as observed by TESS and in the two Strömgren filters in the Appendix (Fig.~\ref{fig:PS_TESS}--\ref{fig:PS_Str_y}) for cycle 19 only. 

The impact of the inclination on the power spectrum becomes more evident in Fig.~\ref{fig:PS_incl_ratios}, where we show the ratios between the power as it would be measured at inclinations of 57$^\circ$ and 0$^\circ$ relative to that obtained by an ecliptic-bound observer.
In agreement with Fig.~\ref{fig:PS_Kepler_TESS} the power on timescales below 200 days decreases with decreasing inclinations, whereas longward of 200 days the power increases with decreasing inclination.
The reason for the increase of the variability is due to several effects. Most noteworthy are the effects of foreshortening and centre-to-limb variations (CLV). In the wavelength regime where \textit{Kepler} operates, the facular contrast (compared to the quiet Sun) is higher at the limb due to limb-darkening, whereas the spot contrast is the strongest at disc centre, as seen by an ecliptic bound observer. With decreasing inclination, the effect of CLV on the facular component is less pronounced and the facular contribution to the brightness variations is increasing (conversely, the effect of the spots is decreasing). While the effect of foreshortening is decreasing with decreasing inclination, it is not enough to compensate for the stronger contrast of the faculae.
For a more detailed discussion see \cite{Shapiro2016}. The distribution of the magnetic features (in particular the spot distribution) is also important, as we discuss in the next section.

\subsection{Disentangling evolution and rotation of magnetic features}\label{disentangle}

\begin{figure*}
\centering
\includegraphics[width=\textwidth]{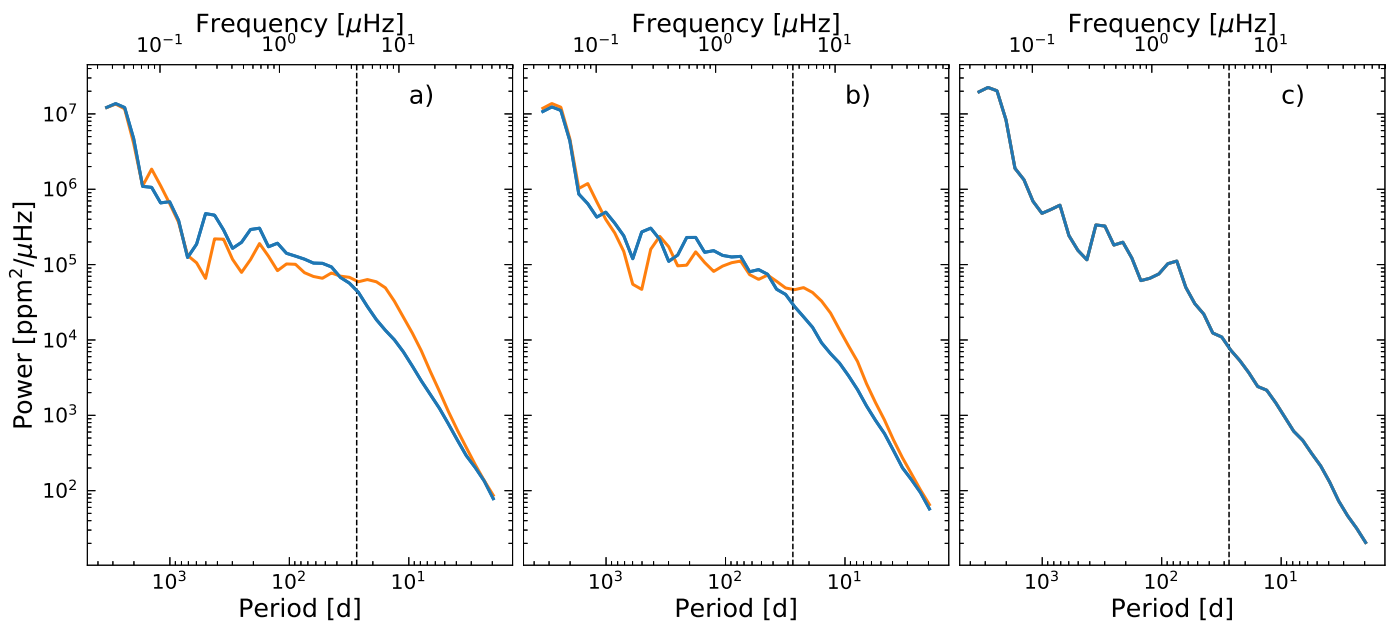}
\caption{Comparison of the power spectra of the solar brightness variations in the  \textit{Kepler} passband, with and without taking the solar rotation into account (orange and blue, respectively). Panel a) shows 90$^\circ$ inclination, b) 57$^\circ$ and c) 0$^\circ$. The vertical dashed black lines indicates the synodic solar rotation period at 27.3 days.} 
\label{fig:PS_effect_nonrot}
\end{figure*}
The solar brightness variability is caused by changes in the solar disc coverage by magnetic features. These changes are in turn due to 
(1) emergence and evolution of magnetic features and (2) the solar rotation,  which causes transits of individual magnetic features across the visible solar disk \citep[see, e.g.][and references therein]{MPS_AA}.  
Our model allows us to pinpoint the contribution of the solar rotation to the solar brightness variability. This can be done by disregarding the free term in Eq.~(\ref{diff_rot}), i.e. by looking at the non-rotating Sun from a fixed direction. We note that by doing this we still preserve the differential rotation term.

In Fig.~\ref{fig:PS_effect_nonrot} we compare the power spectra of solar brightness variations over cycle 22 calculated with and without taking solar rotation into account (orange and blue lines, respectively).
Figure~\ref{fig:PS_effect_nonrot} a shows power spectra as recorded by an ecliptic-bound observer. The solar rotation does not play a big role at timescales below about 4--5 days (the orange and blue lines in Fig.~\ref{fig:PS_effect_nonrot} a are very close to each other). The variability at such timescales is apparently due to the evolution of individual magnetic features. The variability at timescales between 5 days and the solar rotation period is mainly due to the solar rotation. Interestingly, while the rotation itself becomes unimportant at timescales above 
the rotation period the two power spectra are still different up to the timescale of about 4--5 years. This is because the variability of the rotating Sun is determined by the longitudinal-averaged distribution of magnetic features. The variability of 
the non-rotating Sun is given by the distribution seen from a fixed vantage point.
Since the emergence of magnetic features is random over longitude, the two described distributions are the same if averaged over a sufficiently long time interval (so that blue and orange lines almost coincide 
at timescales larger than 4--5 years). At the same time at timescales shorter than 4--5 years the distributions might still be different since they depend on the specific realisation of emergences of magnetic features. Consequently, this part of the power spectrum depends on the 
specific longitudinal location of the vantage point.

Figure~\ref{fig:PS_effect_nonrot} b illustrates the case of 57$^\circ$ inclination, which looks very similar to the case of the ecliptic-bound observer. Fig.~\ref{fig:PS_effect_nonrot} c  represents the view from the observer located over the solar North pole.
Naturally, the solar rotation does not contribute to the brightness variability as it is determined solely by the evolution of the magnetic features and the modulation of their emergence rate over the solar activity cycle. Therefore, the blue and orange curves 
in Fig.~\ref{fig:PS_effect_nonrot} c coincide at all timescales.

Fig.~\ref{fig:PS_effect_nonrot} allows us to better understand the 
origin of the decrease of short-timescale variability  with decreasing inclination as seen in Figs.~\ref{fig:PS_Kepler_TESS}--\ref{fig:PS_incl_ratios}. The emergence of active regions is confined to about $\pm$ 30--40$^\circ$ centred around the equator. Consequently, even though the variability at timescales shorter than 4--5 days is not affected by the solar rotation, it is strongly decreased due to the effect of foreshortening.

\subsection{The full time series}

\begin{figure}
\centering
\includegraphics[width=0.5\textwidth]{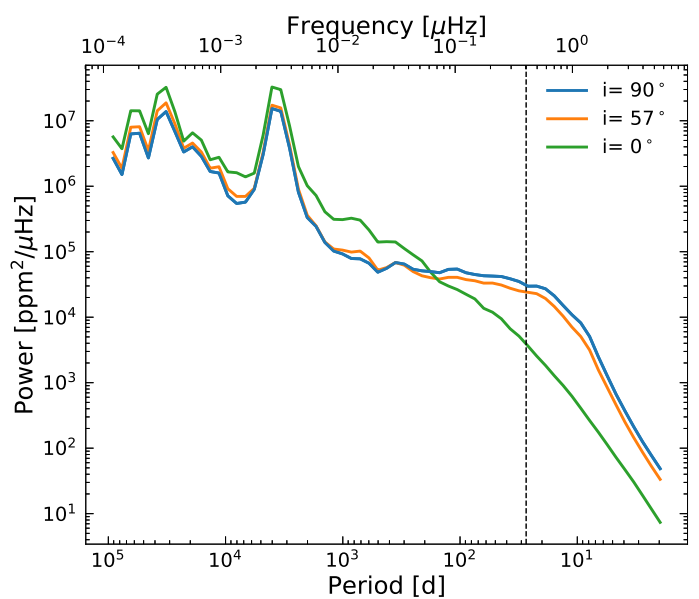}
\caption{Power spectra of the solar brightness variations for the full time series of over 300 years as it would be observed by \textit{Kepler} for different inclinations. The vertical dashed black lines indicates the synodic solar rotation period at 27.3 days.} 
\label{fig:PS_full_series}
\end{figure}

In the previous sections we have limited our analysis to selected individual solar activity cycles. The source term used in the SFTM provides information from 1700 to 2009. We now consider the solar brightness variations for this whole interval, with respect to different inclinations, limiting ourselves to calculating solar brightness variation in the \textit{Kepler} passband, which we present in Fig.~\ref{fig:PS_full_series}. 
The differences in the power spectra between the 90$^\circ$ and 57$^\circ$ degree vantage point are small (shown earlier in the paper), whereas the difference between 90 and 0$^\circ$ is pronounced.
On the timescale above one year, the variability as observed from an inclination of of 0$^\circ$ becomes stronger, due to the stronger facular contribution to the solar brightness variations.

For all inclinations, a pronounced peak at around 10.8 years is visible, which corresponds to the average length of a cycle in our considered sample.
On the rotational timescale, however, no peak is seen and no peaks above or below the rotation period appear. 

\section{Conclusions and outlook}\label{conclusion}
We have employed the Surface Flux Transport model \citep[SFTM in the form of][]{Cameron2010} with the source term from \cite{Jiang2011_1} to obtain the distribution of magnetic flux on the entire solar surface. This distribution was then  converted into surface area coverages of solar magnetic  features and the SATIRE approach was utilised for calculating brightness variations. This allowed modelling the brightness variability of the Sun at different activity levels as it would be seen from any arbitrary vantage point and in  different filter systems.
 
 We have analysed the dependence of the power spectrum of solar brightness variations on the inclination. While the decrease of the inclination leads to an increase of the variability on the timescale of the solar activity cycle the variability decreases at shorter timescales. In particular, it decreases on the timescale of solar rotation. Since the Sun is always seen equator-on, its variability is higher than of another star with the same activity level, but seen from a higher latitude. Consequently the higher variability of solar-like stars cannot be due to the inclinations of their rotation axis alone. The effect of the inclination strengthen the conclusions of \cite{Timo2020} that stars with near-solar fundamental parameters and rotation periods have on average significantly higher variability on the solar rotation timescale than the Sun. 

 Our calculations also indicate that the power spectrum of solar brightness variations does not have a clear peak at the rotation period, not only for the ecliptic-bound observer \citep[see][]{Shapiro2017, Veronika_rot}, but also for the out-of-ecliptic observer. This factor might play an important role in explaining the deficiency of stars with detected near-solar rotation periods \citep[see][]{VanSaders2019, Veronika_rot}.

 Our model also allowed us to decompose the contributions of solar rotation and evolution of magnetic features into solar brightness variability.  In particular, we have shown that the variability on timescales below 5 days is mainly due to the evolution of magnetic features and not due to the solar rotation.

The SFTM model is capable of simulating stars more active than the Sun \citep{Isik2018}, so that we plan to extend the present study to model brightness variations of stars more active than the Sun. By combining it with the results of \cite{witzkeetal2018, Veronika_rot} we also plan to extend it to stars with different fundamental parameters.

\begin{acknowledgements}
The research leading to this paper has received funding from the European Research Council under the European Union’s Horizon 2020 research and innovation program (grant agreement No. 715947). It also got financial support  from the BK21 plus program through the National Research Foundation (NRF) funded by the Ministry of Education of Korea. We would like to thank the International Space Science Institute, Bern, for their support of science team 446 and the resulting helpful discussions.
\end{acknowledgements}

\bibliographystyle{aa}
\bibliography{bib}

\begin{thebibliography}{56}
\expandafter\ifx\csname natexlab\endcsname\relax\def\natexlab#1{#1}\fi

\bibitem[{{Aigrain} {et~al.}(2015){Aigrain}, {Llama}, {Ceillier}, {Chagas},
  {Davenport}, {Garc{\'{\i}}a}, {Hay}, {Lanza}, {McQuillan}, {Mazeh}, {de
  Medeiros}, {Nielsen}, \& {Reinhold}}]{Aigrain2015}
{Aigrain}, S., {Llama}, J., {Ceillier}, T., {et~al.} 2015, \mnras, 450, 3211

\bibitem[{{Angus} {et~al.}(2018){Angus}, {Morton}, {Aigrain}, {Foreman-Mackey},
  \& {Rajpaul}}]{Angus2018}
{Angus}, R., {Morton}, T., {Aigrain}, S., {Foreman-Mackey}, D., \& {Rajpaul},
  V. 2018, \mnras, 474, 2094

\bibitem[{{Baglin} {et~al.}(2006){Baglin}, {Auvergne}, {Boisnard}, {Lam-Trong},
  {Barge}, {Catala}, {Deleuil}, {Michel}, \& {Weiss}}]{COROT}
{Baglin}, A., {Auvergne}, M., {Boisnard}, L., {et~al.} 2006, in COSPAR Meeting,
  Vol.~36, 36th COSPAR Scientific Assembly, 3749

\bibitem[{{Ball} {et~al.}(2012){Ball}, {Unruh}, {Krivova}, {Solanki},
  {Wenzler}, {Mortlock}, \& {Jaffe}}]{Ball2012}
{Ball}, W.~T., {Unruh}, Y.~C., {Krivova}, N.~A., {et~al.} 2012, \aap, 541, A27

\bibitem[{{Baumann} {et~al.}(2006){Baumann}, {Schmitt}, \&
  {Sch{\"u}ssler}}]{Baumann2006}
{Baumann}, I., {Schmitt}, D., \& {Sch{\"u}ssler}, M. 2006, \aap, 446, 307

\bibitem[{{Baumann} {et~al.}(2004){Baumann}, {Schmitt}, {Sch{\"u}ssler}, \&
  {Solanki}}]{Baumann2004}
{Baumann}, I., {Schmitt}, D., {Sch{\"u}ssler}, M., \& {Solanki}, S.~K. 2004,
  \aap, 426, 1075

\bibitem[{{Baumann} \& {Solanki}(2005)}]{Baumann2005}
{Baumann}, I. \& {Solanki}, S.~K. 2005, \aap, 443, 1061

\bibitem[{{Bord{\'e}} {et~al.}(2003){Bord{\'e}}, {Rouan}, \&
  {L{\'e}ger}}]{COROT2}
{Bord{\'e}}, P., {Rouan}, D., \& {L{\'e}ger}, A. 2003, \aap, 405, 1137

\bibitem[{{Borucki} {et~al.}(2010){Borucki}, {Koch}, {Basri}, {Batalha},
  {Brown}, {Caldwell}, {Caldwell}, {Christensen-Dalsgaard}, {Cochran},
  {DeVore}, {Dunham}, {Dupree}, {Gautier}, {Geary}, {Gilliland}, {Gould},
  {Howell}, {Jenkins}, {Kondo}, {Latham}, {Marcy}, {Meibom}, {Kjeldsen},
  {Lissauer}, {Monet}, {Morrison}, {Sasselov}, {Tarter}, {Boss}, {Brownlee},
  {Owen}, {Buzasi}, {Charbonneau}, {Doyle}, {Fortney}, {Ford}, {Holman},
  {Seager}, {Steffen}, {Welsh}, {Rowe}, {Anderson}, {Buchhave}, {Ciardi},
  {Walkowicz}, {Sherry}, {Horch}, {Isaacson}, {Everett}, {Fischer}, {Torres},
  {Johnson}, {Endl}, {MacQueen}, {Bryson}, {Dotson}, {Haas}, {Kolodziejczak},
  {Van Cleve}, {Chandrasekaran}, {Twicken}, {Quintana}, {Clarke}, {Allen},
  {Li}, {Wu}, {Tenenbaum}, {Verner}, {Bruhweiler}, {Barnes}, \&
  {Prsa}}]{KEPLER}
{Borucki}, W.~J., {Koch}, D., {Basri}, G., {et~al.} 2010, Science, 327, 977

\bibitem[{{Cameron} {et~al.}(2010){Cameron}, {Jiang}, {Schmitt}, \&
  {Sch{\"u}ssler}}]{Cameron2010}
{Cameron}, R.~H., {Jiang}, J., {Schmitt}, D., \& {Sch{\"u}ssler}, M. 2010,
  \apj, 719, 264

\bibitem[{{Castelli} \& {Kurucz}(1994)}]{Castelli1994}
{Castelli}, F. \& {Kurucz}, R.~L. 1994, \aap, 281, 817

\bibitem[{{Dasi-Espuig} {et~al.}(2014){Dasi-Espuig}, {Jiang}, {Krivova}, \&
  {Solanki}}]{DasiEspuig2014}
{Dasi-Espuig}, M., {Jiang}, J., {Krivova}, N.~A., \& {Solanki}, S.~K. 2014,
  \aap, 570, A23

\bibitem[{{Dasi-Espuig} {et~al.}(2016){Dasi-Espuig}, {Jiang}, {Krivova},
  {Solanki}, {Unruh}, \& {Yeo}}]{DasiEspuig2016}
{Dasi-Espuig}, M., {Jiang}, J., {Krivova}, N.~A., {et~al.} 2016, \aap, 590, A63

\bibitem[{{Ermolli} {et~al.}(2013){Ermolli}, {Matthes}, {Dudok de Wit},
  {Krivova}, {Tourpali}, {Weber}, {Unruh}, {Gray}, {Langematz}, {Pilewskie},
  {Rozanov}, {Schmutz}, {Shapiro}, {Solanki}, \& {Woods}}]{TOSCA2013}
{Ermolli}, I., {Matthes}, K., {Dudok de Wit}, T., {et~al.} 2013, Atmosph. Chem.
  Phys., 13, 3945

\bibitem[{{Fligge} {et~al.}(2000){Fligge}, {Solanki}, \& {Unruh}}]{Fligge2000}
{Fligge}, M., {Solanki}, S.~K., \& {Unruh}, Y.~C. 2000, \aap, 353, 380

\bibitem[{{Fontenla} {et~al.}(1993){Fontenla}, {Avrett}, \&
  {Loeser}}]{Fontenla1993}
{Fontenla}, J.~M., {Avrett}, E.~H., \& {Loeser}, R. 1993, \apj, 406, 319

\bibitem[{{Fr{\"o}hlich}(2006)}]{Froehlich2006}
{Fr{\"o}hlich}, C. 2006, \ssr, 125, 53

\bibitem[{{Hathaway} \& {Choudhary}(2008)}]{Hathaway2008}
{Hathaway}, D.~H. \& {Choudhary}, D.~P. 2008, \solphys, 250, 269

\bibitem[{{I{\textcommabelow s}{\i}k} {et~al.}(2018){I{\textcommabelow
  s}{\i}k}, {Solanki}, {Krivova}, \& {Shapiro}}]{Isik2018}
{I{\textcommabelow s}{\i}k}, E., {Solanki}, S.~K., {Krivova}, N.~A., \&
  {Shapiro}, A.~I. 2018, \aap, 620, A177

\bibitem[{{Jiang} {et~al.}(2011{\natexlab{a}}){Jiang}, {Cameron}, {Schmitt}, \&
  {Sch{\"u}ssler}}]{Jiang2011_1}
{Jiang}, J., {Cameron}, R.~H., {Schmitt}, D., \& {Sch{\"u}ssler}, M.
  2011{\natexlab{a}}, \aap, 528, A82

\bibitem[{{Jiang} {et~al.}(2011{\natexlab{b}}){Jiang}, {Cameron}, {Schmitt}, \&
  {Sch{\"u}ssler}}]{Jiang2011_2}
{Jiang}, J., {Cameron}, R.~H., {Schmitt}, D., \& {Sch{\"u}ssler}, M.
  2011{\natexlab{b}}, \aap, 528, A83

\bibitem[{Karoff {et~al.}(2018)Karoff, Metcalfe, Ângela R.~G.~Santos, Montet,
  Isaacson, Witzke, Shapiro, Mathur, Davies, Lund, Garcia, Brun, Salabert,
  Avelino, van Saders, Egeland, Cunha, Campante, Chaplin, Krivova, Solanki,
  Stritzinger, \& Knudsen}]{Karoff2018}
Karoff, C., Metcalfe, T.~S., Ângela R.~G.~Santos, {et~al.} 2018, The
  Astrophysical Journal, 852, 46

\bibitem[{{Keppens} \& {Martinez Pillet}(1996)}]{Keppens1996}
{Keppens}, R. \& {Martinez Pillet}, V. 1996, \aap, 316, 229

\bibitem[{{Knaack} {et~al.}(2001){Knaack}, {Fligge}, {Solanki}, \&
  {Unruh}}]{Knaack2001}
{Knaack}, R., {Fligge}, M., {Solanki}, S.~K., \& {Unruh}, Y.~C. 2001, \aap,
  376, 1080

\bibitem[{{Krivova} {et~al.}(2003){Krivova}, {Solanki}, {Fligge}, \&
  {Unruh}}]{Krivova2003}
{Krivova}, N.~A., {Solanki}, S.~K., {Fligge}, M., \& {Unruh}, Y.~C. 2003, \aap,
  399, L1

\bibitem[{{Kurucz}(1992)}]{Kurucz1992}
{Kurucz}, R.~L. 1992, Revista Mexicana de Astronomia y Astrofisica, vol. 23,
  23, 45

\bibitem[{{Lanza} \& {Shkolnik}(2014)}]{Lanza_Shkolnik2014}
{Lanza}, A.~F. \& {Shkolnik}, E.~L. 2014, \mnras, 443, 1451

\bibitem[{{Lockwood} \& {Skiff}(1990)}]{Lockwood1990}
{Lockwood}, G.~W. \& {Skiff}, B.~A. 1990, in NASA Conference Publication, Vol.
  3086, 8--15

\bibitem[{{Lockwood} {et~al.}(2007){Lockwood}, {Skiff}, {Henry}, {Henry},
  {Radick}, {Baliunas}, {Donahue}, \& {Soon}}]{Lockwood2007}
{Lockwood}, G.~W., {Skiff}, B.~A., {Henry}, G.~W., {et~al.} 2007, The
  Astrophysical Journal Supplement Series, 171, 260

\bibitem[{{Martinez Pillet} {et~al.}(1993){Martinez Pillet}, {Moreno-Insertis},
  \& {Vazquez}}]{MartinezPillet1993}
{Martinez Pillet}, V., {Moreno-Insertis}, F., \& {Vazquez}, M. 1993, \aap, 274,
  521

\bibitem[{{McQuillan} {et~al.}(2014){McQuillan}, {Mazeh}, \&
  {Aigrain}}]{McQuillan2014}
{McQuillan}, A., {Mazeh}, T., \& {Aigrain}, S. 2014, The Astrophysical Journal
  Supplement Series, 211, 24

\bibitem[{{Montet} {et~al.}(2017){Montet}, {Tovar}, \&
  {Foreman-Mackey}}]{Montet2017}
{Montet}, B.~T., {Tovar}, G., \& {Foreman-Mackey}, D. 2017, \apj, 851, 116

\bibitem[{{Moreno-Insertis} \& {Vazquez}(1988)}]{MorenoInsertis1988}
{Moreno-Insertis}, F. \& {Vazquez}, M. 1988, \aap, 205, 289

\bibitem[{{Petrovay} \& {van Driel-Gesztelyi}(1997)}]{Petrovay1997}
{Petrovay}, K. \& {van Driel-Gesztelyi}, L. 1997, \solphys, 176, 249

\bibitem[{{Radick} {et~al.}(2018){Radick}, {Lockwood}, {Henry}, {Hall}, \&
  {Pevtsov}}]{Radick2018}
{Radick}, R.~R., {Lockwood}, G.~W., {Henry}, G.~W., {Hall}, J.~C., \&
  {Pevtsov}, A.~A. 2018, \apj, 855, 75

\bibitem[{{Reinhold} {et~al.}(2019){Reinhold}, {Bell}, {Kuszlewicz}, {Hekker},
  \& {Shapiro}}]{Reinhold2019}
{Reinhold}, T., {Bell}, K.~J., {Kuszlewicz}, J., {Hekker}, S., \& {Shapiro},
  A.~I. 2019, \aap, 621, A21

\bibitem[{{Reinhold} {et~al.}(2017){Reinhold}, {Cameron}, \&
  {Gizon}}]{Timo2017}
{Reinhold}, T., {Cameron}, R.~H., \& {Gizon}, L. 2017, \aap, 603, A52

\bibitem[{{Reinhold} {et~al.}(2013){Reinhold}, {Reiners}, \&
  {Basri}}]{Timo2013}
{Reinhold}, T., {Reiners}, A., \& {Basri}, G. 2013, \aap, 560, A4

\bibitem[{{Reinhold} {et~al.}(2020){Reinhold}, {Shapiro}, {Solanki}, {Krivova},
  {Cameron}, \& {Amazo-G\'{o}mez}}]{Timo2020}
{Reinhold}, T., {Shapiro}, A.~I., {Solanki}, S.~K., {et~al.} 2020, Science,
  submitted

\bibitem[{{Ricker} {et~al.}(2014){Ricker}, {Winn}, {Vanderspek}, {Latham},
  {Bakos}, {Bean}, {Berta-Thompson}, {Brown}, {Buchhave}, {Butler}, {Butler},
  {Chaplin}, {Charbonneau}, {Christensen-Dalsgaard}, {Clampin}, {Deming},
  {Doty}, {De Lee}, {Dressing}, {Dunham}, {Endl}, {Fressin}, {Ge}, {Henning},
  {Holman}, {Howard}, {Ida}, {Jenkins}, {Jernigan}, {Johnson}, {Kaltenegger},
  {Kawai}, {Kjeldsen}, {Laughlin}, {Levine}, {Lin}, {Lissauer}, {MacQueen},
  {Marcy}, {McCullough}, {Morton}, {Narita}, {Paegert}, {Palle}, {Pepe},
  {Pepper}, {Quirrenbach}, {Rinehart}, {Sasselov}, {Sato}, {Seager},
  {Sozzetti}, {Stassun}, {Sullivan}, {Szentgyorgyi}, {Torres}, {Udry}, \&
  {Villasenor}}]{TESS}
{Ricker}, G.~R., {Winn}, J.~N., {Vanderspek}, R., {et~al.} 2014, in \procspie,
  Vol. 9143, Space Telescopes and Instrumentation 2014: Optical, Infrared, and
  Millimeter Wave, 914320

\bibitem[{{Schatten}(1993)}]{Schatten1993}
{Schatten}, K.~H. 1993, \jgr, 98, 18

\bibitem[{{Schou} {et~al.}(2012){Schou}, {Scherrer}, {Bush}, {Wachter},
  {Couvidat}, {Rabello-Soares}, {Bogart}, {Hoeksema}, {Liu}, {Duvall}, {Akin},
  {Allard}, {Miles}, {Rairden}, {Shine}, {Tarbell}, {Title}, {Wolfson},
  {Elmore}, {Norton}, \& {Tomczyk}}]{HMI}
{Schou}, J., {Scherrer}, P.~H., {Bush}, R.~I., {et~al.} 2012, \solphys, 275,
  229

\bibitem[{{Shapiro} {et~al.}(2020){Shapiro}, {Amazo-G\'omez}, {Krivova}, \&
  {Solanki}}]{Shapiro2019}
{Shapiro}, A.~I., {Amazo-G\'omez}, E.~M., {Krivova}, N.~A., \& {Solanki}, S.~K.
  2020, A\&A, 633, A32

\bibitem[{{Shapiro} {et~al.}(2017){Shapiro}, {Solanki}, {Krivova}, {Cameron},
  {Yeo}, \& {Schmutz}}]{Shapiro2017}
{Shapiro}, A.~I., {Solanki}, S.~K., {Krivova}, N.~A., {et~al.} 2017, Nature
  Astronomy, 1, 612

\bibitem[{{Shapiro} {et~al.}(2014){Shapiro}, {Solanki}, {Krivova}, {Schmutz},
  {Ball}, {Knaack}, {Rozanov}, \& {Unruh}}]{Shapiro2014}
{Shapiro}, A.~I., {Solanki}, S.~K., {Krivova}, N.~A., {et~al.} 2014, \aap, 569,
  A38

\bibitem[{{Shapiro} {et~al.}(2016){Shapiro}, {Solanki}, {Krivova}, {Yeo}, \&
  {Schmutz}}]{Shapiro2016}
{Shapiro}, A.~I., {Solanki}, S.~K., {Krivova}, N.~A., {Yeo}, K.~L., \&
  {Schmutz}, W.~K. 2016, \aap, 589, A46

\bibitem[{{Snodgrass}(1983)}]{Snodgrass1983}
{Snodgrass}, H.~B. 1983, \apj, 270, 288

\bibitem[{{Solanki} {et~al.}(2013){Solanki}, {Krivova}, \& {Haigh}}]{MPS_AA}
{Solanki}, S.~K., {Krivova}, N.~A., \& {Haigh}, J.~D. 2013, Ann. Rev. Astron.
  Astrophys., 51, 311

\bibitem[{{Unruh} {et~al.}(1999){Unruh}, {Solanki}, \& {Fligge}}]{Unruh1999}
{Unruh}, Y.~C., {Solanki}, S.~K., \& {Fligge}, M. 1999, \aap, 345, 635

\bibitem[{{van Ballegooijen} {et~al.}(1998){van Ballegooijen}, {Cartledge}, \&
  {Priest}}]{vanBallegooijen1998}
{van Ballegooijen}, A.~A., {Cartledge}, N.~P., \& {Priest}, E.~R. 1998, \apj,
  501, 866

\bibitem[{{van Saders} {et~al.}(2019){van Saders}, {Pinsonneault}, \&
  {Barbieri}}]{VanSaders2019}
{van Saders}, J.~L., {Pinsonneault}, M.~H., \& {Barbieri}, M. 2019, \apj, 872,
  128

\bibitem[{{Vieira} {et~al.}(2012){Vieira}, {Norton}, {Dudok de Wit},
  {Kretzschmar}, {Schmidt}, \& {Cheung}}]{Vieira2012}
{Vieira}, L.~E.~A., {Norton}, A., {Dudok de Wit}, T., {et~al.} 2012, \grl, 39,
  L16104

\bibitem[{{Wenzler} {et~al.}(2004){Wenzler}, {Solanki}, {Krivova}, \&
  {Fluri}}]{Wenzler2004}
{Wenzler}, T., {Solanki}, S.~K., {Krivova}, N.~A., \& {Fluri}, D.~M. 2004,
  \aap, 427, 1031

\bibitem[{{Witzke} {et~al.}(2020){Witzke}, {Reinhold}, {Shapiro}, {Krivova}, \&
  {Solanki}}]{Veronika_rot}
{Witzke}, V., {Reinhold}, T., {Shapiro}, A.~I., {Krivova}, N.~A., \& {Solanki},
  S.~K. 2020, arXiv e-prints, arXiv:2001.01934

\bibitem[{{Witzke} {et~al.}(2018){Witzke}, {Shapiro}, {Solanki}, {Krivova}, \&
  {Schmutz}}]{witzkeetal2018}
{Witzke}, V., {Shapiro}, A.~I., {Solanki}, S.~K., {Krivova}, N.~A., \&
  {Schmutz}, W. 2018, \aap, 619, A146

\bibitem[{{Yeo} {et~al.}(2014){Yeo}, {Krivova}, {Solanki}, \&
  {Glassmeier}}]{Yeo2014}
{Yeo}, K.~L., {Krivova}, N.~A., {Solanki}, S.~K., \& {Glassmeier}, K.~H. 2014,
  \aap, 570, A85

\end{thebibliography}

\begin{appendix}
\section{Power spectra of solar brightness variations for TESS and Strömgren \textit{b} and \textit{y}}

\begin{figure}
\centering
\includegraphics[width=0.5\textwidth]{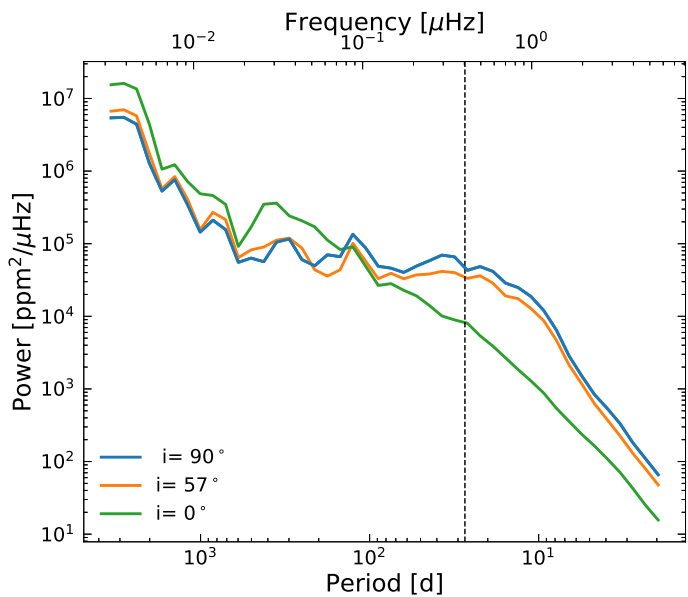}
\caption{Power spectra of the solar brightness variations for the TESS passbands for different inclinations.} 
\label{fig:PS_TESS}
\end{figure}

\begin{figure}
\centering
\includegraphics[width=0.5\textwidth]{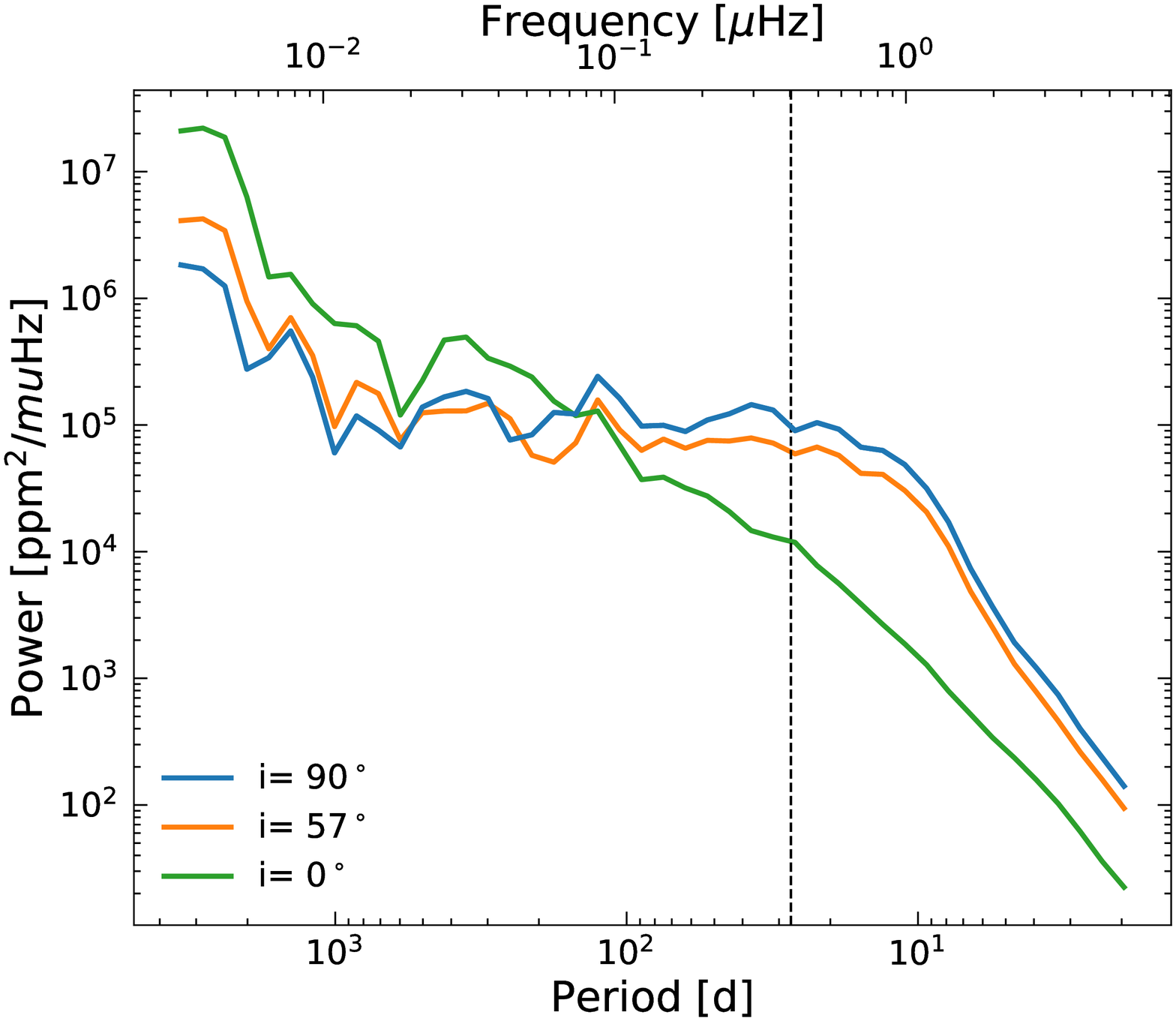}
\caption{Power spectra of the solar brightness variations for Strömgren \textit{b} for different inclinations.} 
\label{fig:PS_Str_b}
\end{figure}

\begin{figure}
\centering
\includegraphics[width=0.5\textwidth]{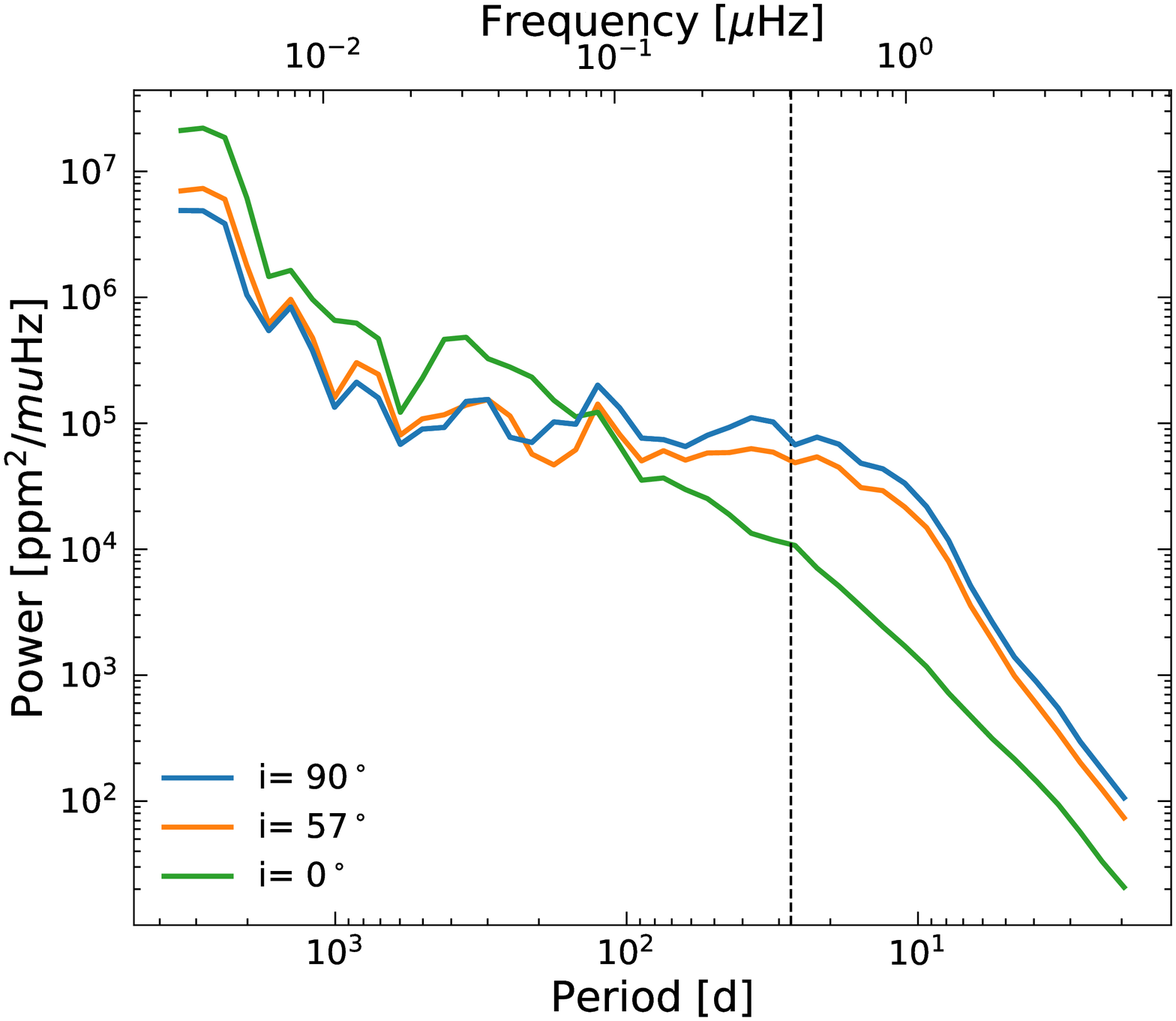}
\caption{Power spectra of the solar brightness variations for  Strömgren \textit{y} for different inclinations.} 
\label{fig:PS_Str_y}
\end{figure}

\end{appendix}

\end{document}